%% file: BAD-1954.tex
\documentclass[twocolumn,showpacs,aps,prl,superscriptaddress]{revtex4}
\usepackage{graphicx}
\usepackage{dcolumn}
\usepackage{epsfig}
\usepackage{amsmath}
\RequirePackage{xspace}

\newcommand{\BaBarPubYear} {08}
\newcommand{\BaBarPubNumber}  {012}
\newcommand{\SLACPubNumber} {13228}

\def\epem  {\ensuremath{e^+e^-}\xspace}
\newcommand\dbline{\noalign{\vskip 0.10truecm\hrule}\noalign{\vskip 2pt}\noalign{\hrule\vskip 0.10truecm}}
\providecommand{\tbline}{\noalign{\vskip 0.05truecm\hrule\vskip0.05truecm}}

\newcommand{\bma}[1]{\boldmath{$#1$}}
\providecommand{\bfemsix}{${\cal B} (10^{-6})$}

\newcommand{\thetaT}{\ensuremath{\theta_{\rm T}}}

\def\ra                 {\ensuremath{\rightarrow}\xspace}

\newcommand{\fetakgamma}{\ensuremath{\eta K \gamma}}
\newcommand{\fetakpgamma}{\ensuremath{\eta K^+ \gamma}}

\newcommand{\fetaggkpgamma}{\ensuremath{\eta_{\gamma \gamma} K^+ \gamma}}

\newcommand{\fetatrepikpgamma}{\ensuremath{\eta_{3\pi} K^+ \gamma}}

\newcommand{\fetakzgamma}{\ensuremath{\eta K^{0} \gamma}}

\newcommand{\fetaggkzgamma}{\ensuremath{\eta_{\gamma \gamma} K^{0} \gamma}}

\newcommand{\fetatrepikzgamma}{\ensuremath{\eta_{3\pi} K^{0} \gamma}}

\providecommand{\BetaKGamma}{\mbox{$B \rightarrow \eta K \gamma $}}
\providecommand{\BetaKpGamma}{\mbox{$B^+ \rightarrow \eta K^+ \gamma $}}

\newcommand{\deltat}{\mbox{$\Delta t$}}
\newcommand{\dt}{\mbox{$\Delta t$}}
\newcommand{\ttag}{\mbox{$t_{\rm tag}$}}
\newcommand{\deltamd}{\mbox{$\Delta m_d$}}
\newcommand{\bflav}{\ensuremath{B_{\rm flav}}}

\newcommand{\etaKzg}{\mbox{$B^0\ra  \eta K^0  \gamma$}}
\newcommand{\etaKzsg}{\mbox{$B^0\ra  \eta \KS  \gamma$}}
\newcommand{\etaKpg}{\mbox{$B^+\ra  \eta K^+  \gamma$}}

\newcommand{\psfile}[3][]{ 
  \begin{center}
    \setlength{\epsfxsize}{#3\linewidth}\leavevmode
    \def\noOpt{}\def\testit{#1}\ifx\testit\noOpt%
      \epsfbox{#2}%
    \else%
      \epsfbox[#1]{#2}%
    \fi
  \end{center}
}
\def\gaga  {\ensuremath{\gamma\gamma}\xspace} 
\def\pip   {\ensuremath{\pi^+}\xspace}
\def\pim   {\ensuremath{\pi^-}\xspace}
\def\piz   {\ensuremath{\pi^0}\xspace}

\newcommand{\etagg}{\ensuremath{\eta_{\gaga}}}
\newcommand{\etappp}{\mbox{$\eta_{3\pi}$}}

\def\KS    {\ensuremath{K^0_{\scriptscriptstyle S}}}
\def\Bu      {\ensuremath{B^+}}
\def\Bub     {\ensuremath{B^-}}
\def\Bbar    {\overline{B}{}}

\def\Bzb     {\ensuremath{\Bbar^0}}
\def\Bz      {\ensuremath{B^0}}
\def\BpBm    {\ensuremath{\Bu  \Bub}}
\def\BzBzb   {\ensuremath{\Bz  \Bzb}}

\newcommand{\DE}{\ensuremath{\Delta E}}
\def\mes{\mbox{$m_{\rm ES}$}}
\newcommand{\calB}{\mbox{${\cal B}$}}
\newcommand{\acp}{\ensuremath{{\cal A}_{ch}}}

\def\babar{{\em B}{\footnotesize\em A}{\em B}{\footnotesize\em AR}}
\def\pep2{PEP-II}

\newcommand{\UfourS}{\mbox{$\Upsilon(4S)$}}

\newcommand\etal{{\it et al.}}

\newcommand{\gevcc}{\mbox{$\textrm{GeV}/c^2$}} 
\newcommand{\mevcc}{\mbox{$\textrm{MeV}/c^2$}} 
 
\newcommand{\gev}{\mbox{$\textrm{GeV}$}}

\newcommand{\jprlBase}  [1]     {Phys.\ Rev.\ Lett. \xspace}
\newcommand{\npbps}     [1]  {{Nucl.\ Phys.\ B~Proc.\ Suppl.\ {\bf #1}}}
\newcommand{\jprl}      [1]    {\jprlBase\  ~{\bf #1}}
\newcommand{\jprBase}        {Phys.\ Rev.\ }
\newcommand{\jprd}      [1]  {\jprBase\ D~{\bf #1}}
\newcommand{\plBase}   [1]         {Phys.\ Lett. \xspace}
\newcommand{\plb}      [1]    {\plBase\  B~{\bf #1}}
\newcommand{\nimBaseA}       {Nucl.\ Instr.\ Meth.\ }
\newcommand{\nima}      [1]  {\nimBaseA~A~{\bf #1}}
\newcommand{\zpBase}         {Z.\ Phys.}
\newcommand{\zpc}       [1]  {\zpBase\ C~{\bf #1}}
\newcommand{\npBase}         {Nucl.\ Phys.\ }
\newcommand{\npb}       [1]  {\npBase\ B~{\bf #1}}

\newcommand{\jrmp}      [1]  {{Rev.\ Mod.\ Phys.\ {\bf #1}}}

\def\BB      {\ensuremath{B\Bbar}\xspace} 
\def\CP                {\ensuremath{C\!P}\xspace}
\newcommand{\half}{\mbox{${1\over2}$}}
\newcommand{\pvec}{{\bf p}}
\def\qqbar{\mbox{$q\bar q\ $}}

\newcommand{\signf}{$\cal S$ ($\sigma$)}

\newcommand{\eff}{$\epsilon$}

  \newcommand{\BretaKzg}{\mbox{$\calB(\etaKzg)$}}
  \newcommand{\RetaKzg}{\ensuremath{7.1^{+2.1}_{-2.0}\pm 0.4}}       
  \newcommand{\setaKzg}{\ensuremath{3.9}}                    

  \newcommand{\SetaKzsg}{\ensuremath{-0.18^{+0.49}_{-0.46}\pm 0.12}} 
  \newcommand{\CetaKzsg}{\ensuremath{-0.32^{+0.40}_{-0.39}\pm 0.07}} 

  \newcommand{\BretaKpg}{\mbox{$\calB(\etaKpg)$}}
  \newcommand{\RetaKpg}{\ensuremath{7.7 \pm 1.0 \pm 0.4}}       
  \newcommand{\setaKpg}{\ensuremath{8.0}}                    

\def\figurebox#1#2#3{%
    \def\arg{#3}%
    \ifx\arg\empty
    {\hfill\vbox{\hsize#2\hrule\hbox to #2{\vrule\hfill\vbox to
          #1{\hsize#2\vfill}\vrule}\hrule}\hfill}%
    \else
    {\hfill\epsfbox{#3}\hfill}%
    \fi}

\begin{document}

\preprint{\babar-PUB-\BaBarPubYear/\BaBarPubNumber} 
\preprint{SLAC-PUB-\SLACPubNumber} 

\begin{flushleft}
~\\
\end{flushleft}

\begin{flushright}
~\\
\babar-PUB-\BaBarPubYear/\BaBarPubNumber \\
SLAC-PUB-\SLACPubNumber \\
\end{flushright}

\title{\large  \bf\boldmath  Branching Fractions and
  \CP-Violating Asymmetries in 
  Radiative  $B$ Decays to \fetakgamma }

\input authors_mar2008.tex

\begin{abstract}
We present  measurements of the  \CP-violation parameters
$S$ and $C$ for the radiative decay \etaKzsg; for  \BetaKGamma\  we also measure the branching fractions 
and for \BetaKpGamma\  the time-integrated charge asymmetry
\acp.
 The data, collected with the \babar\ detector
at the Stanford Linear Accelerator Center, represent $465 \times 10^6$
 \BB\ pairs produced in $e^+e^-$ annihilation.
The results are
$S = \SetaKzsg$,
$C = \CetaKzsg$,
$\BretaKzg = (\RetaKzg) \times 10^{-6}$,
$\BretaKpg = (\RetaKpg) \times 10^{-6}$, and $\acp =
(-9.0^{+10.4}_{-9.8} \pm 1.4) \times 10^{-2}$.
The first error quoted is statistical and the second systematic.
\end{abstract}

\pacs{13.25.Hw, 12.15.Hh, 11.30.Er}

\maketitle

Radiative $B$ meson decays have long been recognized  as
a sensitive probe to test the standard model (SM) and to look 
for new physics (NP)~\cite{Hou, Hurth}. 
In the SM, flavor-changing neutral current processes, such as 
$b \rightarrow s \gamma$, proceed via radiative loop  diagrams.
The loop diagrams may also contain  
new  heavy particles, and therefore are sensitive to NP.
The measured
branching fractions of inclusive $b\rightarrow s \gamma$ 
and exclusive radiative $B$ decays are in agreement with SM
predictions~\cite{Hurth,ThPred,PDG2006}. Recent estimates 
of the branching fraction of  the inclusive  $b \rightarrow s \gamma$
decay are affected by a  theoretical uncertainties as large 
as the experimental ones \cite{PrevisioniTeoriche}. 
 
In the SM the photon polarization in radiative decays is dominantly  
left (right) handed for $b$ ($\bar{b}$) decays, resulting  in the suppression
 of mixing-induced~\cite{Gronau} \CP\ asymmetries. Because we do  not measure the photon helicity,
we sum the decay rates for the left-handed and right-handed helicity states. 
Observation of significant  \CP-violation in these radiative decay
modes would  provide a clear sign of NP~\cite{Atwood}. 
We search also for direct \CP\ asymmetry in charged $B$ decays, 
measuring the charge asymmetry 
 $\acp \equiv (\Gamma^--\Gamma^+)/(\Gamma^-+\Gamma^+)$, where $\Gamma$
is the partial decay width of the $B$ meson, and the superscript
corresponds to its charge. This asymmetry in the SM  is expected to be very 
small~\cite{Grueb}.

In this letter, we present the first measurement of the mixing-induced 
\CP\ violation in the decay mode \etaKzg. 
Branching fractions for the decay modes \etaKzg\ and \etaKpg\ \cite{ChargeCon}
and time-integrated  charge-asymmetry for \etaKpg\ have been measured 
previously
by the  Belle~\cite{BELLE} and  \babar~\cite{Previous} Collaborations.
We update our previous measurements with a data sample that is twice as 
large.

The results presented here are based on data collected
with the \babar\ detector~\cite{BABARNIM}
at the PEP-II asymmetric-energy $e^+e^-$ collider~\cite{pep}
located at the Stanford Linear Accelerator Center.
We use an integrated
luminosity of 423~fb$^{-1}$, corresponding to 
$465 \pm 5$ million \BB\ pairs, recorded at the $\Upsilon (4S)$ 
resonance (at a center-of-mass energy of $\sqrt{s}=10.58\ \gev$).

Description of the \babar\ detector and of the reconstruction of charged and 
neutral particles can be found elsewhere~\cite{Aux}. The $B$ decay daughter 
candidates are reconstructed through their decays
$\piz\ra\gaga$, $\eta\ra\gaga$ (\etagg), and $\eta\ra\pip\pim\piz$
(\etappp). Reconstruction and selection criteria of charged and neutral
mesons, and primary photons and the study  of continuum and \BB\ 
backgrounds are described  in our previous 
paper \cite{Previous}. 

A $B$ meson candidate is reconstructed by combining  an $\eta$
candidate, a charged or neutral kaon and a primary photon
candidate. It is characterized kinematically by the energy-substituted
mass $\mes \equiv \sqrt{(s/2 + \pvec_0\cdot \pvec_B)^2/E_0^2 - \pvec_B^2}$ and
energy difference $\DE \equiv E_B^*-\half\sqrt{s}$, where the
subscripts $0$ and 
$B$ refer to the initial \UfourS\ and to the $B$ candidate in the
lab-frame, respectively, 
and the asterisk denotes the \UfourS\ rest frame. We require $5.25 < \mes <
5.29$ \gevcc\ and $|\DE|<0.2$ \gev.

From a candidate \BB\ pair we reconstruct a \Bz\  decaying into
 $\eta \KS \gamma$ ($B_{\rm rec}$).  We also reconstruct the
decay point of the other 
$B$ meson ($B_{\rm tag}$) and identify its flavor.
The difference $\deltat \equiv t_{\rm rec} - \ttag$
of the proper decay times $t_{\rm rec}$ and $\ttag$ of the
reconstructed and tag $B$ mesons,  
respectively, is obtained from the measured distance between the
$B_{\rm rec}$
and  $B_{\rm tag}$ decay vertices and from the boost ($\beta \gamma =0.56$) of 
the \epem system. The \deltat\ distribution~\cite{kspi0gamma} is given by:
\begin{eqnarray}
  F(\dt) &=& 
        \frac{e^{-\left|\deltat\right|/\tau}}{4\tau} [1 \mp\Delta w \pm
                                                   \nonumber\\
   &&\hspace{-3em}(1-2w)\left( S\sin(\deltamd\deltat) -
C\cos(\deltamd\deltat)\right)].\label{eq:FCPdef}
\end{eqnarray}
The upper (lower) sign denotes a decay accompanied by a \Bz (\Bzb) tag,
$\tau$ is the mean $\Bz$ lifetime, $\deltamd$ is the mixing
frequency, and the mistag parameters $w$ and
$\Delta w$ are the average and difference, respectively, of the probabilities
that a true $\Bz$\ is incorrectly tagged as a $\Bzb$\ or vice versa.
In the flavor tagging algorithm \cite{s2b} there are   six mutually exclusive tagging
categories of 
different response purities and untagged events with no tagging informations.
Tagging and \dt\ informations are used for the measurement of the
\CP-violation parameters  $S$ and $C$ in the decay mode \etaKzg.  
 
We use the same technique developed for $B^0\ra \piz \KS \gamma$
decays~\cite{kspi0gamma} to reconstruct the $B^0\ra \etagg \KS \gamma$
decay point, using the knowledge of the \KS\ trajectory and the average 
interaction point in a geometric fit. 
The extraction of $\deltat$  has been extensively validated in
data~\cite{kspi0prl04} and 
in a full detector simulation. In about 70\% of the selected events the
\dt\ resolution is sufficient for the
time-dependent \CP-violation measurement.
For the remaining events the $\deltat$ information is not used.
For both $\etagg \KS \gamma$ and  $\etappp \KS \gamma$ modes we use
the events which satisfy the requirements $|\deltat|<20$~ps and
$\sigma_{\Delta t}<2.5$~ps, where $\sigma_{\Delta t}$ is the per-event
error on \deltat. 

We obtain signal event yields and \CP-violation parameters from
unbinned extended maximum-likelihood (ML) 
fits. We indicate with $j$ the
species of event: signal, \qqbar\ continuum background, \BB\ peaking 
background ($B\!P$), and \BB\ non-peaking background ($B\!N\!P$).
 The input observables are \mes, \DE, the output of a Neural
Network ($N\!N$), the $\eta$ invariant mass $m_{\eta}$, and \deltat.
The $N\!N$ combines four variables: the absolute values 
of the cosines of the  polar angles with respect to the  
beam axis in the \UfourS\ frame
of the $B$ candidate momentum and the $B$ thrust axis, the ratio of
the second and zeroth Fox-Wolfram moments~\cite{FoxWol}, and the
absolute value of the cosine of the angle  
\thetaT\  between the thrust axis of the $B$ candidate and that of the rest 
of the tracks and neutral clusters in the event, calculated in the \UfourS\ frame.

For each species $j$ and tagging  category       
$c$, we define a total probability density function (PDF) for event $i$ as
\begin{eqnarray}
{\cal P}_{j,c}^i &\equiv & {\cal P}_j ( \mes^i ) \cdot {\cal  P}_j ( \DE^i )
\cdot { \cal P}_j( N\!N^i ) \cdot \nonumber \\ 
& & { \cal P}_j( m_{\eta}^i ) \cdot  
{ \cal  P}_j (\deltat^i, \sigma_{\Delta t}^i;c)\,.
\end{eqnarray}
The factored form of the PDF is a good approximation since correlations
between input observables are small.  
With $n_{j}$ defined to be the number of events of the species $j$
and $f_{j,c}$ the fraction of events of species $j$ for each category $c$,
we write the extended likelihood function for all events belonging to
category $c$ as 
\begin{eqnarray}
{\cal L}_c &=& \exp{\Big(-\sum_{j} n_{j,c}\Big)}
           \prod_i^{N_c} (n_{\rm sig}f_{{\rm sig},c}{\cal P}_{{\rm
               sig},c}^{i}\nonumber \\
                  &&\hspace{-1em}+n_{q\bar{q}} f_{q\bar{q},c}{\cal P}_{q\bar{q}}^{i}
                   +n_{B\!N\!P}f_{B\!N\!P,c}{\cal P}_{B\!N\!P}^{i}
                   \nonumber \\
                   &&\hspace{-1em} +n_{B\!P}f_{B\!P,c}{\cal P}_{B\!P}^{i}),
\end{eqnarray}
where $n_{j,c}$ is the yield of events of species $j$ found by the fitter 
in category $c$ and $N_c$ the number of events of category $c$ in the sample.
We fix $f_{{\rm sig},c}$, $f_{B\!N\!P,c}$, and  $f_{B\!P,c}$ to
$f_{\bflav,c}$, the values measured with a large sample of
$B$-decays to fully reconstructed flavor eigenstates
(\bflav)~\cite{Resol}. 
The total likelihood function ${\cal L}_d$ for decay mode $d$ is given as the
product over the seven tagging categories.  Finally, when combining
decay modes we form the grand likelihood ${\cal L}=\prod{\cal L}_d$. 

The PDF ${ \cal P}_{\rm sig} (\dt,\, \sigma_{\Delta t}; c)$, for each category
$c$, is the  
convolution of $F(\dt;\, c)$ (Eq.\ \ref{eq:FCPdef}) with the
signal resolution function (sum of three Gaussians) determined from the
\bflav\ sample.
The other PDF forms are: the sum of two Gaussians for ${\cal P}_{\rm
sig}(\mes)$, ${\cal P}_{\rm sig}(\DE)$, and ${\cal P}_{\rm sig}(m_{\eta})$;
the sum of three Gaussians for  
${\cal P}_{\qqbar}(\dt)$, ${\cal P}_{B\!N\!P}(\dt)$, and ${\cal P}_{B\!P}(\dt)$; 
a non-parametric step function for ${\cal P}_j(N\!N)$~\cite{Aaron}; a linear
dependence for ${\cal P}_{\qqbar}(\DE)$, ${\cal P}_{B\!N\!P}(\DE)$, and
${\cal P}_{B\!P}(\DE)$; a first-order polynomial plus a Gaussian for
${\cal P}_{\qqbar}(m_{\eta})$, ${\cal P}_{B\!N\!P}(m_{\eta})$, and 
${\cal P}_{B\!P}(m_{\eta})$; 
and for ${\cal 
P}_{\qqbar}(\mes)$, ${\cal P}_{B\!N\!P}(\mes)$, and
${\cal P}_{B\!P}(\mes)$,  the function
$x\sqrt{1-x^2}\exp{\left[-\xi(1-x^2)\right]}$, 
with $x\equiv2\mes/\sqrt{s}$~\cite{argus}, where for the $B\!P$ 
PDFs we add a 
Gaussian. We allow \qqbar\ background PDF
parameters  to vary in the fit.

We determine the PDF parameters from Monte Carlo (MC) simulation for the
signal and \BB\ backgrounds, while using 
   sideband data ($5.25 < \mes\
<5.27$ \gevcc; $0.1<|\DE |<0.2$ \gev )  
to model the PDFs of  continuum  background. 
Large control samples of $B$ decays to charmed final states with   similar 
topology and a smearing procedure applied to photons during the event
reconstruction are used to verify the simulated resolutions in \mes\
and \DE.
Where the control data samples reveal differences from the MC  in mass  
resolution, we shift or scale the resolution used in the likelihood
fits. The largest shift in  \mes\ is $0.6$ \mevcc. 
Any bias in the fit
is determined from a
large set of simulated 
experiments in which the \qqbar\ and $B\!N\!P$ backgrounds are
generated from the PDFs, and 
into which we have embedded  
the expected number of $B\!P$  and signal events 
chosen  randomly from fully simulated MC samples. 

We compute the branching fractions and charge asymmetry from fits made
without \dt\ or flavor tagging.  
The free parameters in the fit are: the signal, \qqbar,
$B\!N\!P$ and $B\!P$ background
yields; the bin weights of the step function for ${\cal
  P}_{\qqbar}(N\!N)$; the slopes of ${\cal 
  P}_{\qqbar}(\DE)$ and ${\cal 
  P}_{\qqbar}(m_{\eta})$; $\xi$; 
and for charged modes the signal and background \acp. 
We apply the same procedure  to extract $S$ and
$C$. In this case we add in the fit 
the \dt\ variable and the flavor-tagging information.
 As free parameters we have
also $S$, $C$, and the parameters of the ${\cal P}_{\qqbar}(\dt)$ PDF.

\begin{table*}[htp]
\caption{
Number of events $N$  in the sample, corrected signal yield,  detection
efficiency  \eff, daughter branching fraction product $\prod\calB_i$,
significance \signf\ (including systematic uncertainties), and measured branching
fraction \calB\ with statistical error for each decay mode. For the
combined measurements we give  \signf\ and the  branching fraction
with statistical and systematic uncertainty. 
For the neutral mode we give the $S$ and $C$ parameters for each decay
mode and for their combination.
For the charged 
modes we also give the measured  signal charge asymmetry \acp. 
}
\label{tab:results}
\begin{tabular}{lcccccccccc}
\dbline
Mode&\quad $N$ \quad  &\quad Yield \quad  & \eff\ (\%)  &$\prod\calB_i$(\%)  
& \signf  &\quad \bfemsix \quad &
\acp\ ($10^{-2}$) & \quad $S$ \quad & \quad $C$ \quad \\
\tbline
~~\fetaggkzgamma    & $3690$& $58^{+19}_{-18}$ &$12$
&$13.6$&$3.3$&$7.4^{+2.5}_{-2.3}$&  & $-0.04 \pm 0.62$ & $-0.24 \pm 0.44$ \\
~~\fetatrepikzgamma &$2282$  & $24^{+13}_{-12}$   &$10$
&$7.8$&$2.1$&$6.6^{+3.6}_{-3.2}$& & $-0.45\pm0.81$& $-0.71\pm0.87$\\
\bma{\fetakzgamma}& & &   &  &{\bma \setaKzg}& {\bma \RetaKzg} & & {\bma \SetaKzsg}  & {\bma \CetaKzsg}  &
\\
\tbline
~~\fetaggkpgamma &  $11620$ & $266^{+37}_{-36}$  &$19$&$39.4$&$6.5$&$7.8^{+1.1}_{-1.0}$&$-4 \pm 12$ \\ 
~~\fetatrepikpgamma &  $10738$& $111^{+26}_{-24}$   &$14$&$22.4$&$4.5$&$7.4^{+1.7}_{-1.6}$&$-24 \pm 20$ \\
\bma{\fetakpgamma} &  & & &    &{\bma \setaKpg} & {\bma \RetaKpg } &${\bf -9.0^{+10.4}_{-9.8} \pm 1.4}$\\
\dbline
\end{tabular}
\vspace{-5mm}
\end{table*}

Table \ref{tab:results} lists the results of the fits.
The corrected signal yield is the fitted yield minus the fit bias which 
is in the range $2-4\%$. 
The efficiency is calculated as the ratio of the number of signal MC
events entering 
 the ML fit to the total generated.  
We compute the branching fractions from the corrected signal 
yields, reconstruction efficiencies, 
daughter branching fractions, and the number of produced $B$ mesons. We
assume that the branching fractions of the \UfourS\ to \BpBm\ and \BzBzb\ are 
each equal to 50\%.  
We combine results from different channels by adding their
likelihood functions,  taking into  account
 the correlated and uncorrelated systematic errors.

The statistical error on the signal yield, $S$, $C$ and the signal
charge asymmetry  is taken as the change in    
the central value when the quantity $-2\ln{\cal L}$ increases by one 
unit from its minimum value. The significance \signf\ is  the square root 
of the difference between the value of $-2\ln{\cal L}$ (with systematic 
uncertainties included) for zero signal and the value at its minimum.

\begin{figure}[tb]
\includegraphics[angle=0,scale=0.4335]{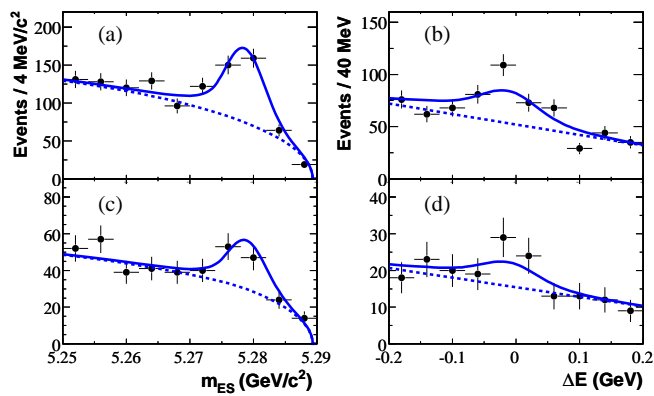}
\vspace{-0.7cm}
 \caption{\label{fig:projections}
 The $B$ candidate \mes\ and \DE\ projections (see text) for \fetakpgamma\ (a, b),
 \fetakzgamma\ (c, d).
 Points with error bars (statistical only) represent the data, the solid
line the full fit function, and the dashed line its background component.  }
\vspace{-.2cm}
\end{figure}

\begin{figure}[!htb]
  \begin{center}
   \includegraphics[scale=0.3]{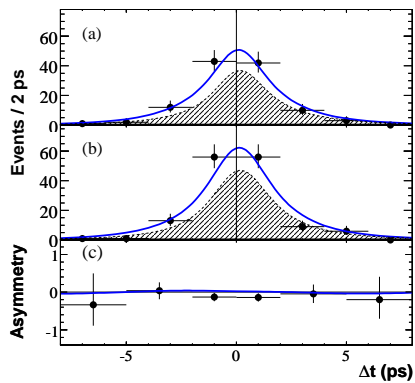}
\end{center}
  \vspace*{-0.5cm}
 \caption{Projections (see text) onto $\Delta t$ of the data (points with error bars),
fit function (solid line), and background function
(dashed line), for (a) \Bz\ and (b) \Bzb\ tagged events, and (c) the
raw asymmetry   $(N_{\Bz}-N_{\Bzb})/(N_{\Bz}+N_{\Bzb})$ between \Bz\ and \Bzb\ tags.}
  \label{fig:DeltaTProj}
\end{figure}

Figure~\ref{fig:projections} shows, as representative fits, 
the  projections onto \mes\ and \DE\ while Fig.~\ref{fig:DeltaTProj}
shows the projections onto  $\Delta t$ and  the raw
asymmetry between \Bz\ and \Bzb\ tags. In these projections
a subset of the data is used for which the signal likelihood
(computed without the variable plotted) exceeds a threshold that optimizes the sensitivity.

\begin{figure}[t]
\resizebox{\columnwidth}{!}{
\begin{tabular}{cc}
\includegraphics[]{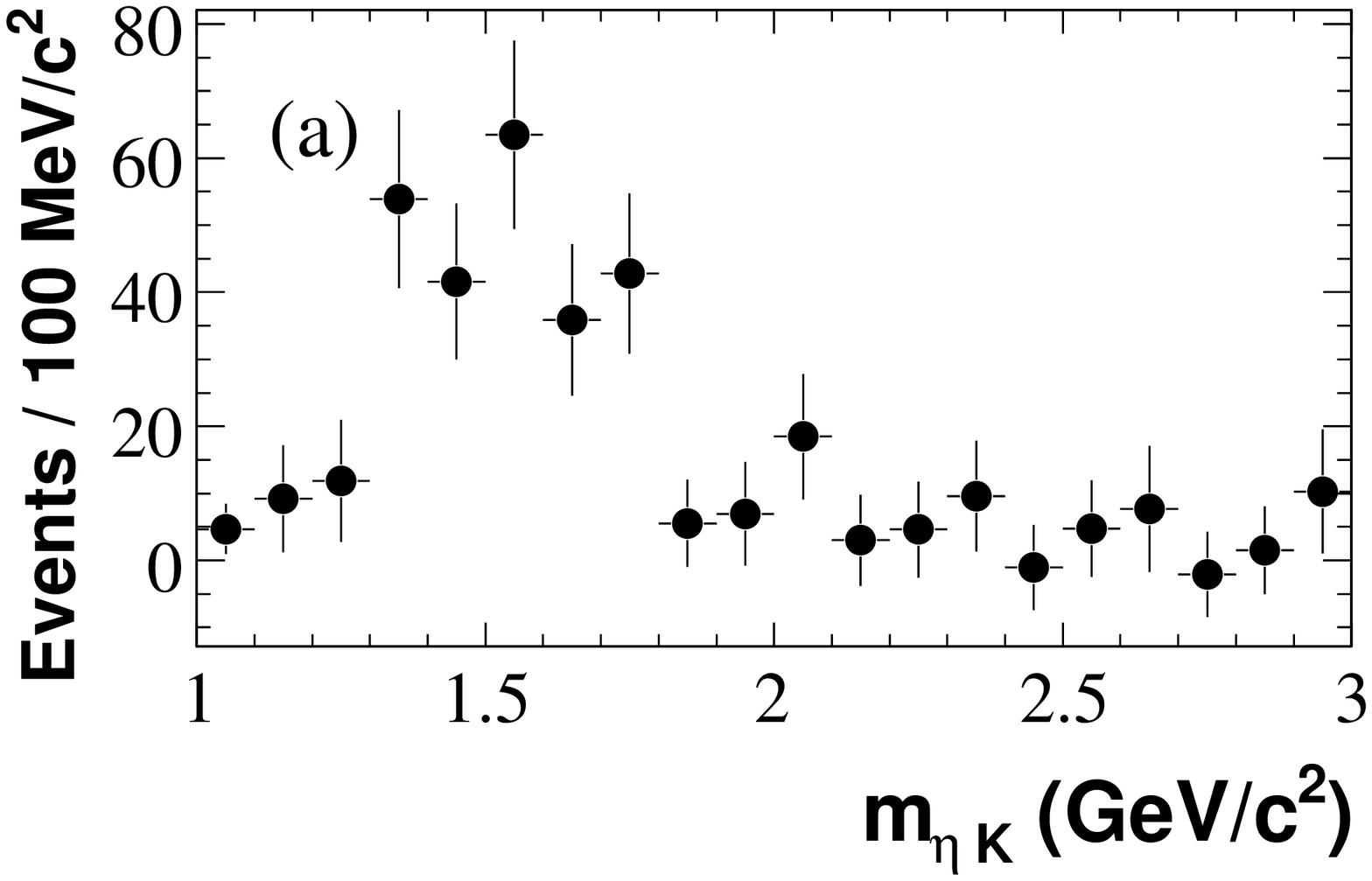} &
\includegraphics[]{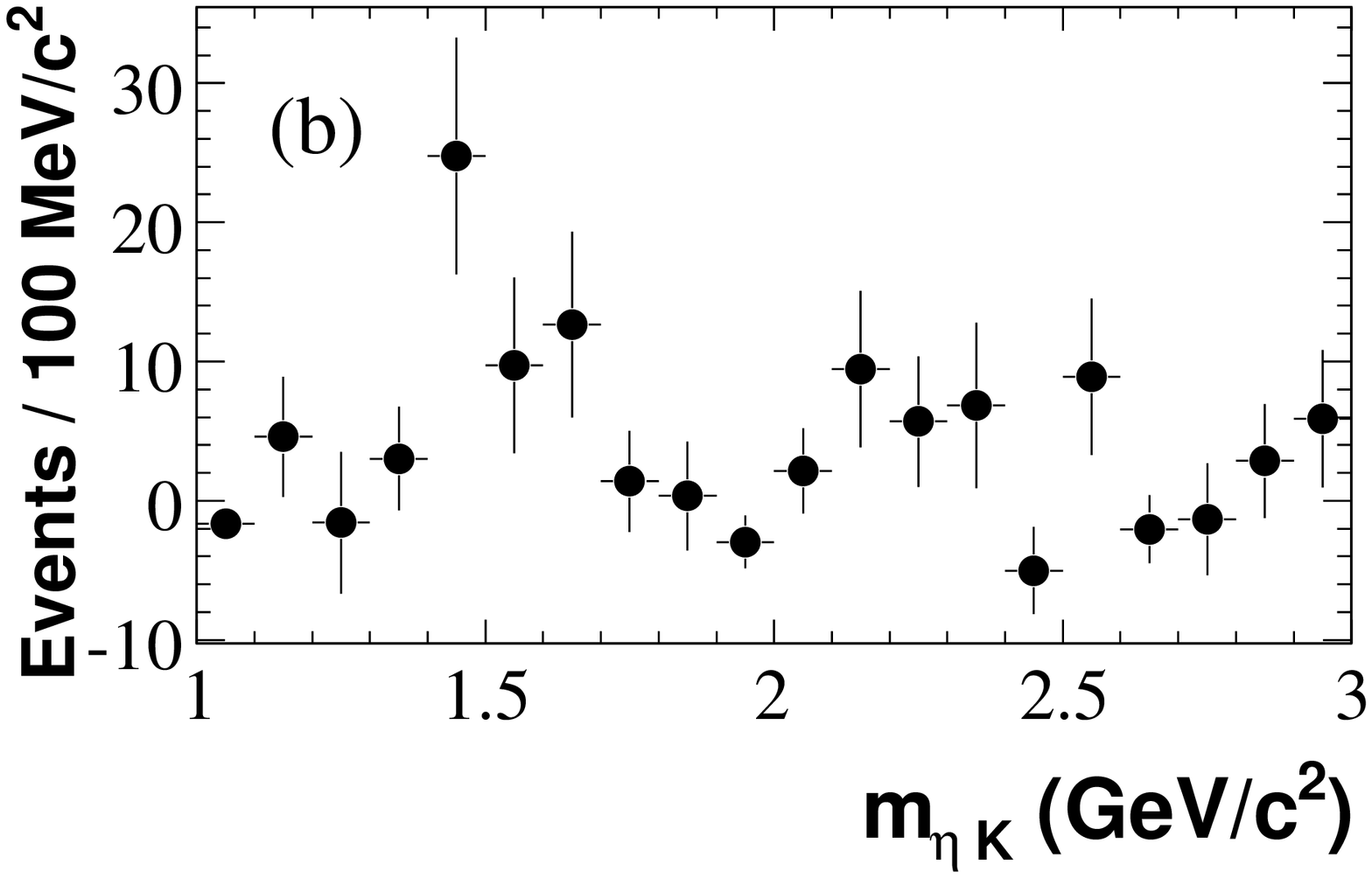} 
\end{tabular}
}
\caption{Plot of $\eta K$ invariant mass for signal using the
  weighting technique described in the text for the combined sub-decay
  modes: (a) \etaKpg,  (b) \etaKzg. Errors are statistical only.}
\label{fig:sPlots}
\end{figure}

Figure~\ref{fig:sPlots} shows the distribution of the $\eta K$
invariant mass for signal events obtained by 
the event-weighting technique (sPlot) described in Ref.~\cite{sPlots}.
We use the covariance matrix and PDFs
from the ML fit  to determine a probability for each signal event.
  The resulting distributions (points with errors) 
are normalized to the signal yield. This mass distribution is useful
to compare with theoretical predictions for radiative decays.

The main sources of systematic uncertainties for the time-dependent measurements
come from the variation of the signal PDF shape parameters within their
errors ($0.08$ for $S$, $0.04$ for $C$), and  from \BB\ backgrounds
($0.09$ for $S$, $0.06$ for $C$).
Other minor sources are SVT alignment, beam spot position and size, 
and interference between the CKM-suppressed $\bar{b}\to\bar{u} c\bar{d}$ 
amplitude   
and the favored $b\to c\bar{u}d$ amplitude for some tag-side $B$ decays
\cite{dcsd}. 
The \bflav\ sample is used  to determine the errors associated with the signal \dt\
resolutions, tagging efficiencies, and mistag rates. Published
measurements \cite{PDG2006} for $\tau$ and \deltamd\ 
are used to determine the errors associated with them.  
Summing all systematic errors in quadrature, we obtain $\pm 0.12$ for $S$
and $\pm 0.07$ for $C$.

The main sources of systematic uncertainties  for the branching fraction measurements
include uncertainties in the
 PDF  parameterization  and ML fit bias.    For the
signal, the uncertainties in PDF parameters are estimated by comparing
MC and data in control samples.  Varying the 
signal PDF parameters within these errors, we estimate yield
uncertainties of 3--23 events, depending on the mode.  The uncertainty
(1--3 events)
from fit bias is taken as half the correction
itself. 
Systematic uncertainties due to lack of knowledge of the primary photon
spectrum are estimated to be in the range 2--3\% depending on the decay mode. 
Uncertainties in our knowledge of the efficiency, found from auxiliary
studies \cite{Aux}, include $0.4\%\times N_t$ and $1.8\%\times N_\gamma$, where
$N_t$ and $N_\gamma$ are the numbers of tracks and photons, respectively,
in the $B$ candidate. There is a  systematic error of 2.1\% in the
efficiency of \KS\ reconstruction. 
  The uncertainty in the total number of \BB\ pairs in the
data sample is 1.1\%.  Published data~\cite{PDG2006}\ provide the
uncertainties in the $B$ daughter branching fraction products (0.7--1.8\%).

A systematic uncertainty of 0.014 is assigned  to \acp. This uncertainty  is estimated  from 
studies with signal MC events and data control samples and from 
calculation of the asymmetry due to particles interacting in the 
detector.

In conclusion,  we measure the time-dependent 
\CP\ violation parameters in the decay mode $\Bz \ra \eta
\KS \gamma$:  $S = \SetaKzsg$ and $C = \CetaKzsg$.  
These results are consistent with  no \CP-violation in
this mode. We also measure the branching fractions, in units of 
$10^{-6}$, \BretaKzg = \RetaKzg\ and
\BretaKpg = \RetaKpg, in agreement with the results from Belle~\cite{BELLE} and the previous 
\babar\ results~\cite{Previous}. 
The measured charge asymmetry in the decay \etaKpg\  is consistent
with zero. Its confidence interval at 90\%  confidence  
level is  [$-0.25,0.08$].

\input acknow_PRL

\end{document}

%% file: authors_mar2008.tex
%
\author{B.~Aubert}
\author{M.~Bona}
\author{Y.~Karyotakis}
\author{J.~P.~Lees}
\author{V.~Poireau}
\author{E.~Prencipe}
\author{X.~Prudent}
\author{V.~Tisserand}
\affiliation{Laboratoire de Physique des Particules, IN2P3/CNRS et Universit\'e de Savoie, F-74941 Annecy-Le-Vieux, France }
\author{J.~Garra~Tico}
\author{E.~Grauges}
\affiliation{Universitat de Barcelona, Facultat de Fisica, Departament ECM, E-08028 Barcelona, Spain }
% ...
\collaboration{The \babar\ Collaboration}
\noaffiliation

%% file: acknow_PRL.tex
We are grateful for the excellent luminosity and machine conditions
provided by our \pep2\ colleagues, 
and for the substantial dedicated effort from
the computing organizations that support \babar.
The collaborating institutions wish to thank 
SLAC for its support and kind hospitality. 
This work is supported by
DOE
and NSF (USA),
NSERC (Canada),
CEA and
CNRS-IN2P3
(France),
BMBF and DFG
(Germany),
INFN (Italy),
FOM (The Netherlands),
NFR (Norway),
MES (Russia),
MEC (Spain), and
STFC (United Kingdom). 
Individuals have received support from the
Marie Curie EIF (European Union) and
the A.~P.~Sloan Foundation.

%% file: BAD-1954.bbl
\begin{thebibliography}{99}
\bibitem{ChargeCon}
Charge-conjugate modes are implied throughout.
\end{thebibliography}
